

\input amstex.tex
\documentstyle{amsppt}
\input amsppt1.tex
\nologo
\magnification=1200

\NoBlackBoxes

\define\op#1{\operatorname{#1}}

\topmatter
\author  A. Kazarnovski-Krol               \endauthor
\title Value of generalized hypergeometric function at unity \endtitle
\address{
  Department of Mathematics
  Rutgers University
  New Brunswick, NJ 08854, USA}

\vskip 10mm
\vskip 10mm

\abstract{
Value of generalized hypergeometric function at
a special point is calculated. More precisely, value of certain
multiple integral over vanishing cycle (all arguments collapse to unity)
is calculated. The answer is expressed in terms of
$\Gamma$-functions. The constant is relevant to the part of $\rho$
in the Gindikin-Karpelevich formula for c-function of Harish-Chandra.
Calculation is an adaptation of classical calculations of Gelfand and Naimark
(1950) to the Heckman-Opdam hypergeometric functions in the case of root system
 of type $A_{n-1}$.
 }
\endtopmatter

\document
\head {0 . Introduction}\endhead

In this paper value of certain multiple integral over vanishing cycle
(all arguments collapse to unity)
is calculated. This is a normalization constant. It is related to
the part of $\rho$ in the Gindikin-Karpelevich formula for
$c$-function of Harish-Chandra.

The cycle is a
distinguished one in the theory of zonal spherical functions. It
appeared for the first time in [1] and its role is clarified in [21].

Technically our calculations is an adaptation of formulas of [1]
(elliptic coordinates)
for the case of generic $k$, i.e. for the Heckman-Opdam hypergeometric
functions of type $A_{n-1}$.

The integral may be regarded as a variation on a theme of Selberg
integral
[19],
and recently integrals of this type have drawn much attention because
of applications to conformal field theory.

The paper has continuation in [20,21].

\vskip 5mm
\head { 1. Theorem}\endhead
Let $z_l,\; l=1,...,n;\; t_{i,j},\;i=1,...,j,\;j=1,...,n-1$ be the set of
real variables. It is convenient to organize these variables in the form
of a pattern cf. fig.1.

\midinsert
$$
\matrix
z_{1}&&z_{2}&&\ldots&&\ldots&&z_{n}\\\\
&t_{1,n-1}&&t_{2,n-1}&&\ldots&&t_{n-1,n-1}&\\\\
&&\ldots&&\ldots&&\ldots&&\\\\
&&&t_{1,2}&&t_{2,2}&&&\\\\
&&&&t_{1,1}&&&&
\endmatrix
$$
\botcaption{Figure 1}
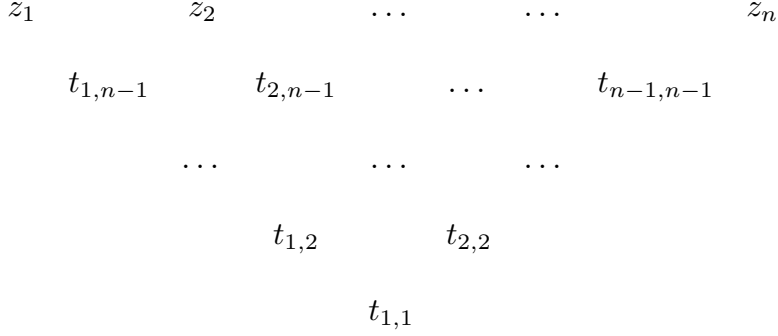
 Variables organized in a pattern
\endcaption
\endinsert

\vskip 0.5cm

\remark{Remark 1.1 } This type of variables reflects the flag
structure ($i$th row corresponds to $i$-dimensional plane in a flag)
,
cf.[1,2]. One should notice that this type of variables is used in [10]
for the isomorphism of Heckman-Opdam hypergeometric system with  a
particular case of trigonometric version of
Knizhnik-Zamolodchikov equation.
\endremark

\definition{Definition 1.2} Consider the following multivalued form
$\omega_{\Delta}:$

$$
\align
\omega_{\Delta}:= &\prod_{i=1}^{n} z_i^{a_i}
 \prod_{i_1 > i_2}{(z_{i_1}-z_{i_2})^{1-2k}}\\
 &\times\prod_{i<l}{(z_l-t_{i,{n-1}})}^{k-1}\prod_{i \ge
l}{(t_{i,{n-1}}-z_l)}^{k-1}  \\
 &\times\prod_{j=1}^{n-1} \{
 \prod_{i_1 \ge i_2} ( t_{i_{1},j} - t_{i_2,{j+1}} )^{k-1}
\prod_{i_1 > i_2} ( t_{i_{1},j+1 } - t_{i_2,j} )^{k-1}\\
 &\times\prod_{i_1>i_2} {(t_{i_1,j}-t_{i_2,j})^{2-2k} }\\
 &\times\prod_{i=1}^{j} {t_{ij}^{a_{ij} }} \}\quad
 {dt_{11} dt_{12} dt_{22} \ldots dt_{n-1,n-1}  }
\endalign
$$

\enddefinition
\smallskip

\remark{Remark 1.3} For the applications one should let
$a_{ij}= {\lambda_{n-j+1}-
 \lambda_{n-j} -k}$ and
$a_i={\lambda_1 +{k {(n-1)}\over 2}}$,

but we formulate the theorem in this more general form
(with $a_{ij}$ and $a_i$ unspecified).
\endremark
\smallskip

\definition{Definition 1.4}
Set
$$
\Phi(z_1,\ldots,z_n)= \int_{\Delta} \omega_{\Delta}
, $$
where $\omega_{\Delta}$ is defined above and ,
assuming that $z_1, z_2, \ldots , z_{n}$ are real and satisfy
$0 < z_1 <z_2 < \ldots <z_{n}$,
define cycle $\Delta$ by the following inequalities:
$t_{i,j+1} \le t_{ij} \le t_{i+1, j+1}$ and
$z_i \le t_{i,n-1} \le z_{i+1}$ .

We assume that phases
of the factors of the form $\omega_{\Delta}$ are equal to zero
provided $k$ , $a_{ij}$, and $a_i$ are real.
\enddefinition

\smallskip

\proclaim{Theorem 1.5}
The limit of $\Phi(z_1,\ldots,z_n)$ as all $z_i$
approach $1$, while $z_1<z_2<\ldots<z_n$, is equal to:
$$
{\Phi(1,\ldots,1):={\lim\limits_{all\; {z_i}\rightarrow
1}}\Phi(z_1,\ldots,z_n)}=
 {
 {\Gamma(k)\Gamma(k)^2\ldots\Gamma(k)^n}
 \over
 {\Gamma(k)\Gamma(2k) \ldots\Gamma(nk)}
 }\quad .
$$
\endproclaim

\subhead{Proof} \endsubhead
  Following the classical work of I.M. Gelfand and
M.A. Naimark cf. [1]
 ,  let
$$
\tau_{ij}=
{
 {\prod\limits_{i_1=1}^{j-1}(t_{i_1,j-1}-t_{ij})}
 \over
 {\prod\limits_{i_1\ne i}(t_{i_1,j}-t_{ij})}
}\;\; ,
$$
$i=1,\ldots,j$
$j=1,\ldots,n-1$.
Note that ${\sum\limits_{i=1}^j}\tau_{ij}=1$, and
$$
{
 {D(\tau_{1,j},\ldots,\tau_{j-1,j})}
 \over
 {D(t_{1,j-1},\ldots,t_{j-1,j-1})}
}
=
{
 {{\prod\limits_{1\le i<k\le j-1}}(t_{i,j-1}-t_{k,j-1})}
 \over
 {{\prod\limits_{1\le i<p\le j}}(t_{ij}-t_{pj})}
}
$$
[see \cite{1} for the details].
Let also
$$
\tau_{in}={
           {{\prod\limits_{i=1}^{n-1}}(t_{i_1,n-1}-z_i)}
           \over
           {{\prod\limits_{i_1\ne i}}(z_{i_1}-z_i)}
          }
\;\;\;\;\;\;\;\;\;\;\;  i=1,\ldots,n.
$$
One has ${\sum\limits_{i=1}^n}\tau_{in}=1$ and
$$
{
 {D(\tau_{1,n},\ldots,\tau_{n-1,n})}
 \over
 {D(t_{1,n-1},\ldots,t_{n-1,n-1})}
}
=
{
 {{\prod\limits_{1\le i<k\le n-1}}(t_{i,n-1}-t_{k,n-1})}
 \over
 {{\prod\limits_{1\le i<p\le n}}(z_i-z_p)}
}
$$
In the variables $\tau_{ij}\;\;\; i=1,\ldots,j-1,\;\; j=1,\ldots,n$
integral for $\Phi(z_1,\ldots,z_n)$ is written as:
$$
\align
&\Phi(z_1,\ldots,z_n)
=
\prod_{i=1}^n{z_i^{a_i}}\int{\prod\limits_{j=1}^{n-1} \;
              \prod\limits_{i=1}^j
               }
    (t_{ij})^{a_{ij}}\\
&\times
{
{\prod\limits_{j=1}^n}
 (
  (\tau_{1j}\tau_{2j}\ldots\tau_{j-1,j})
  (1-\tau_{1j}-\ldots -\tau_{j-1,j})
 )^{k-1}
}
d\tau_{12}d\tau_{13}d\tau_{23}\ldots d\tau_{1n}\ldots
d\tau_{n-1,n}\;\; .
\endalign
$$
Integration is taken over
$$
\tau_{ij}>0,\;\;\;\;\;\;\;\; {\sum\limits_{i=1}^{j-1}}\tau_{ij}<1.
$$
As all $z_i$, $i=1,\ldots,n$ approach $1$, all $t_{ij}$ also approach
$1$, so
$$
\align
\Phi(1,\ldots,1)=&
\int 1 \times
{
 {\prod\limits_{j=1}^n}
                            (
                              (\tau_{1j}\tau_{2j}\ldots\tau_{j-1,j})
                              (1-\tau_{1j}-\ldots -\tau_{j-1,j})
                            )^{k-1}
}\\
&\times d\tau_{12}d\tau_{13}d\tau_{23}\ldots d\tau_{1n}\ldots
d\tau_{n-1,n}.
\endalign
$$
So using Dirichlet's formula one gets
$$
{\Phi(1,\ldots,1)}=
{
 {\Gamma(k)\Gamma(k)^2\ldots\Gamma(k)^n}
 \over
 {\Gamma(k)\Gamma(2k) \ldots\Gamma(nk)}
}
$$
\remark{ Remark 1.6} The same calculation as in theorem 1.5. shows
that
 $\Phi(z_1,z_2,\ldots,z_n)$ is equal to the same constant as in
theorem
provided that all  $a_{ij}=0$ and $a_i=0$. Then this constant is
interpreted as the volume of a maximal compact subgroup
(under certain normalizations)cf. [1,13,14] in the case of generic $k$, i.e.
when
there is no group and no subgroup at all.
The fact that this constant does not depend on $z_i$ also implies the
monodromy properties of cycle $\Delta$, cf. [21].
\endremark
\subhead Acknowledgments \endsubhead I would like to express my gratitude to
I. M. Gelfand for many stimulating discussions on the representation
theory and the theory of hypergeometric functions. I would like to
thank S. Shatashvili for helpful discussions.

\enddocument